\documentclass[sigconf]{acmart}
\usepackage{tabularx}
\usepackage{array}
\newcolumntype{C}{>{\centering\arraybackslash}X}
\usepackage{tipa}
\usepackage{enumitem}

\AtBeginDocument{%
  }

\copyrightyear{2025}
\acmYear{2025}
\setcopyright{acmlicensed}
\acmConference[MM '25]{Proceedings of the 33rd ACM International Conference on Multimedia}{October 27--31, 2025}{Dublin, Ireland}
\acmBooktitle{Proceedings of the 33rd ACM International Conference on Multimedia (MM '25), October 27--31, 2025, Dublin, Ireland}
\acmDOI{10.1145/3746027.3758148}
\acmISBN{979-8-4007-2035-2/2025/10}

\begin{document}

\title{Singing Timbre Popularity Assessment Based on Multimodal Large Foundation Model}

% 净化作者

% 正常作者

\author{Zihao Wang}
\affiliation{%
  \institution{Zhejiang University} % [cite: 3]
  \city{Hangzhou}
  \country{China}
}
\affiliation{%
  \institution{Carnegie Mellon University} % [cite: 3]
  \city{Pittsburgh}
  \country{United States}
}
\email{carlwang@zju.edu.cn}

\author{Ruibin Yuan}
\affiliation{%
  \institution{Hong Kong University of Science and Technology}
  \city{Hongkong}
  \country{China}
}
\email{a43992899@gmail.com}

\author{Ziqi Geng}
\affiliation{%
  \institution{University of California, Berkeley}
  \city{Berkeley}
  \country{United States}
}
\email{ag40804@berkeley.edu}

\author{Hengjia Li}
\affiliation{%
  \institution{Zhejiang University} % [cite: 3]
  \city{Hangzhou}
  \country{China}
}
\affiliation{%
  \institution{Carnegie Mellon University} % [cite: 3]
  \city{Pittsburgh}
  \country{United States}
}
\email{lihengjia98@gmail.com}

\author{Xingwei Qu}
\affiliation{%
  \institution{University of Manchester}
  \city{Manchester}
  \country{United Kingdom}
}
\email{xingwei.qu@postgrad.manchester.ac.uk}

\author{Xinyi Li}
\affiliation{
  \institution{Zhejiang University}
  \city{Hangzhou}
  \country{China}
}
\email{xinyili@zju.edu.cn}

\author{Songye Chen}
\affiliation{%
  \institution{Mei KTV}
  \city{Beijing}
  \country{China}
}
\email{3156018231@qq.com}

\author{Haoying Fu}
\affiliation{
  \institution{Mei KTV} 
  \city{Beijing}
  \country{China}
}
\email{326452438@qq.com}

\author{Roger B. Dannenberg}
\affiliation{
  \institution{Carnegie Mellon University} 
  \city{Pittsburgh}
  \country{United States}
}
\email{rbd@cs.cmu.edu}

\author{Kejun Zhang}
\authornote{Corresponding author}
\affiliation{
  \institution{Zhejiang University}
  \city{Hangzhou}
  \country{China}
}
\affiliation{
  \institution{Innovation Center of Yangtze River Delta, Zhejiang University}
  \city{Hangzhou}
  \country{China}
}
\email{zhangkejun@zju.edu.cn}

\renewcommand{\shortauthors}{Zihao Wang et al.}

\begin{abstract}
Automated singing assessment is crucial for education, entertainment, and talent discovery. However, existing systems are hindered by two fundamental limitations: first, their reliance on reference tracks (e.g., the original song), which stifles creative expression, and second, their simplification of complex vocal performances into a single, often non-diagnostic score based on pitch and rhythm. This paradigm fails to capture the nuanced, multifaceted attributes that define expert-level singing. Echoing the recent shift in other AI domains from discriminative to descriptive evaluation, we advocate for a new paradigm in singing assessment. This paper aims to build a complete ecosystem for reference-free, multi-dimensional, and descriptive singing assessment. First, we construct Sing-MD, a large-scale, multi-dimensional singing dataset annotated by experts across four core dimensions: breath control, timbre quality, emotional expression, and vocal technique. Analysis of this dataset reveals a key finding: significant annotation inconsistencies among experts, which challenges the validity of traditional accuracy-based evaluation metrics. Second, standard Multimodal Large Language Models (MLLMs) are unable to analyze full-length songs on resource-constrained, consumer-grade hardware due to memory limitations. This challenge leads to a ``human label-audio input mismatch'' problem and results in poor performance. To address this issue, we designed VocalVerse, an efficient hybrid architecture. It leverages a lightweight acoustic encoder and specialized modules to process the entire song, thereby learning global performance features, modeling long-term dependencies, and ultimately overcoming this limitation.
Third, to address the shortcomings of automated metrics, we establish a new evaluation benchmark—H-TPR (Human-in-the-loop Tiered Perceptual Ranking)—which evaluates a model's ability to generate perceptually valid performance rankings, rather than predicting a noisy ``ground-truth'' score.
Our comprehensive experiments show that on the H-TPR benchmark, our VocalVerse framework can effectively learn and distinguish singing quality across different dimensions, thereby creating perceptually valid quality rankings and significantly outperforming existing baselines. 
Furthermore, our framework for multi-dimensional scoring and descriptive feedback generation has been successfully commercialized and deployed at scale, demonstrating its significant real-world impact and practical value.
\end{abstract}





\maketitle
\section{INTRODUCTION}
Singing is a core mode of human expression, central to cultural heritage, entertainment, and personal development. The proliferation of music creation and consumption platforms, from karaoke apps to professional singing competitions, has spurred a huge demand for objective, insightful, and scalable feedback on vocal performance. The rapid development of generative models for speech and music has also created new possibilities for automated content analysis~\cite{copet2024simple, dhariwal2020jukebox, 2022songdriverarxiv, wang2024remastarxiv, 10222365, wang2024muchinarxiv, 2024muditarxiv}. However, the field of Automatic Singing Quality Assessment (ASQA) has long been dominated by a paradigm fundamentally inconsistent with the nature of artistic expression.

The first limitation of traditional ASQA systems is their \textbf{reference dependency}~\cite{Tsai2011Karaoke, Gupta2018TechnicalFramework, Tsai2012AutomaticEvaluationFeatures}. These systems compare a user's pitch and rhythm against a predefined template of the original song. While this is useful for basic karaoke scoring, this approach penalizes creative interpretations, stylistic variations, and original compositions, effectively rejecting any performance that deviates from the preset path.

To overcome this, reference-free methods have emerged, but they often fall into a second trap: \textbf{dimensionality singularity}. Most systems distill a rich, complex vocal performance into a single, ambiguous score~\cite{Zhang2019BiDense, Huang2020SpectralFeatures, Gupta2020TwinNeural, Ju2024EndToEnd}. A score of ``7 out of 10'' offers no diagnostic value. Does it reflect poor breath control, a lack of emotional connection, or flawed vocal technique? This is in stark contrast to how human experts, such as vocal coaches, evaluate singing. They provide multi-faceted, descriptive feedback, targeting specific areas for improvement like breath support, timbre control~\cite{Ju2023ImprovingASSE, Sun2023TGCritic}, emotional delivery, and technical skill.

Finally, the field suffers from an \textbf{evaluation fallacy}. The standard practice is to train models to predict a ``ground-truth'' score provided by human annotators and to measure success using metrics like Mean Absolute Error (MAE) or accuracy. However, as we will demonstrate, this assumes the existence of a single, objective truth in an inherently subjective domain. We find that even trained experts exhibit significant disagreement when scoring, making the pursuit of perfect accuracy a chase after a noisy, ill-defined target.

Inspired by recent paradigm shifts in other AI domains like emotion recognition~\cite{lian2025affectgpt}—which are moving from simple classification to rich, descriptive understanding using Multimodal Large Language Models (MLLMs)—we believe the time is ripe for a similar revolution in singing assessment. We argue that the goal of singing assessment should not be just to give a score, but to provide understanding.

In this paper, we design a holistic framework that fundamentally redefines reference-free singing assessment.\footnote{Available at: \url{https://github.com/CarlWangChina/Singing-Aesthetic-Assessment}} We treat it as a multi-dimensional, descriptive task, building a complete ecosystem that includes a novel dataset, a dedicated model, and a more meaningful evaluation benchmark. Our contributions are as follows:

\begin{itemize}[leftmargin=*]
    \item \textbf{A New Dataset and a Core Finding:} We introduce \textbf{Sing-MD}, a large-scale dataset of user-generated singing performances, annotated by vocal experts across four key dimensions: \textit{breath control, timbre quality, emotional expression,} and \textit{vocal technique}. Crucially, our analysis is the first to quantitatively reveal the significant inter-annotator disagreement among experts, exposing the limitations of traditional evaluation metrics.
    \item \textbf{An Efficient Full-Song Model Architecture:} To address the practical challenge of processing full-length song audio on consumer hardware—where standard MLLMs face the critical \textbf{inconsistency between manual annotations and clip-based audio inputs} due to VRAM limitations—we developed \textbf{VocalVerse}. This novel hybrid architecture uses lightweight acoustic encoders to process the entire song, enabling it to learn the global context that is lost in clip-based approaches and achieving a balance between performance and efficiency.
    \item \textbf{A New Human-Centric Evaluation Paradigm:} Recognizing the flaws of automated metrics in this subjective task, we design the \textbf{H-TPR (Human-in-the-loop Tiered Perceptual Ranking)} benchmark. H-TPR no longer measures a model's ability to match a noisy score, but rather assesses its ability to generate a perceptually consistent ranking of performances, which is more aligned with real-world applications like competition judging and personalized feedback.
    \item \textbf{Real-World Deployment and Impact:} We demonstrate the practical viability of our framework through its successful commercialization. The system has been deployed in over 800 karaoke venues, providing AI-driven, ``vocal coach''-style feedback to over 200,000 users each month, showcasing its significant real-world utility and impact.
\end{itemize}

\section{RELATED WORK}
Our work is positioned at the intersection of Automatic Singing Quality Assessment (ASQA) and the emerging paradigm of descriptive evaluation powered by Multimodal Large Language Models (MLLMs). This section reviews the evolution of these fields to contextualize our contributions.

\subsection{Evolution of Singing Assessment Paradigms}
Early ASQA research was predominantly reference-based, focusing on objectively quantifiable metrics. These systems evaluated pitch accuracy by comparing a sung melody to a ground-truth score~\cite{Tsai2011Karaoke, Gupta2018TechnicalFramework} and assessed rhythmic fidelity by analyzing timing deviations~\cite{Tsai2012AutomaticEvaluationFeatures}. Other measurable features included pitch stability, vibrato characteristics~\cite{Nakano2006AutomaticEvaluation}, and basic acoustic parameters derived from signal processing~\cite{Omori1996SingingPowerRatio}. While foundational, the reliance on reference scores limited their application to creative performances, spurring the development of reference-free ASQA, which acknowledges that human listeners can evaluate singing quality without a familiar reference~\cite{Nakano2006Subjective, Nakano2006AutomaticEvaluation}.

The advent of deep learning marked a significant paradigm shift, enabling models to learn complex patterns directly from audio data. A variety of neural network architectures have been employed. Convolutional Neural Networks (CNNs) have been favored for their ability to extract local spectro-temporal features from representations like Mel-spectrograms and Constant-Q Transforms (CQT)~\cite{Zhang2019BiDense, Huang2020SpectralFeatures}. To model temporal dependencies, Recurrent Neural Networks (RNNs) with GRU or LSTM cells became standard~\cite{Gupta2020TwinNeural, Huang2020SpectralFeatures}, often combined with CNNs into hybrid CRNN structures~\cite{Huang2020SpectralFeatures}. To provide global context, models have been augmented with auxiliary features such as pitch and rhythm histograms~\cite{Gupta2020TwinNeural, Huang2020SpectralFeatures, Ju2023ImprovingASSE, Gupta2021Rhythm}. More recently, attention mechanisms~\cite{Vaswani2017Attention} were introduced to allow models to focus on the most salient parts of a performance~\cite{Ju2023ImprovingASSE, Ju2024EndToEnd}, and multi-task learning has been used to predict multiple attributes, such as pitch and an overall score, simultaneously~\cite{Li2021TrainingExplainable}.

\subsection{Challenges in Assessing Nuanced Vocal Quality}
While deep learning architectures have grown in sophistication, assessing nuanced and subjective aspects of singing beyond basic technical accuracy remains a significant challenge. Timbre, the unique ``color'' of a voice, is crucial for the aesthetic experience~\cite{McAdams2009PerceptionTimbre}, but its assessment is complicated by its subjectivity and its deep entanglement with other performance factors~\cite{Wapnick1997ExpertConsensus}.

Early attempts to incorporate such nuances often utilized speaker embedding techniques like X-vectors~\cite{Snyder2018Xvectors} or specialized timbre models like CROSS~\cite{Lee2019CROSS} as features for a general quality score~\cite{Ju2023ImprovingASSE, Ju2024EndToEnd}. Some works, like TG-Critic~\cite{Sun2023TGCritic}, used such embeddings to guide the assessment process. However, a critical issue with these approaches is their reliance on datasets annotated with labels derived from indirect user signals like ``likes'' or ``comments.'' These signals are poor proxies for intrinsic quality, as they are heavily influenced by non-vocal factors like song popularity or lyrical content. To improve model robustness against acoustic variations, techniques like perturbation-based data augmentation~\cite{Chen2025TrainingBetterEmbedding}, mixing strategies for accompanied singing~\cite{Ju2024EndToEnd}, and Singing Voice Separation (SVS) as a preprocessing step~\cite{Li2022VocEmb4SVS} have been explored. Despite these efforts, the output of most ASQA systems remains a singular, holistic score, failing to provide the specific, dimensional feedback characteristic of human expert evaluation.

\subsection{The Shift Towards Descriptive Evaluation with MLLMs}
The limitations of single-score paradigms have led to a necessary shift in perspective. The advent of MLLMs has catalyzed a paradigm shift in many AI fields from discriminative tasks to generative or descriptive ones. A prime example is in multimodal emotion recognition, where works like AffectGPT~\cite{lian2025affectgpt} argue that pre-defined emotion categories are insufficient and instead leverage MLLMs to generate rich, natural language descriptions of emotional states. This trend is mirrored in music AI, where the focus is shifting towards aligning models with human intent, exemplified by work on emotion-based arrangement~\cite{remast}, melody pre-training~\cite{wu2023melodyglm}, and the development of user-aligned models~\cite{mudit-acl} and new comprehensive benchmarks~\cite{wang2024muchin}.

This broader movement highlights a shift towards AI systems that provide understanding and explanation. In contrast to the slow evolution in assessment, AI music generation models now tackle highly complex tasks such as real-time accompaniment~\cite{2022songdriver}, full-song synthesis~\cite{yuan2025yue}, and zero-shot singing voice conversion~\cite{dai2025everyone, wang2024samoyearxiv}, highlighting a significant gap between generative capability and assessment sophistication. Our work applies the descriptive philosophy to bridge this gap in singing assessment. To our knowledge, we are the first to systematically frame it as a descriptive task for MLLMs, tackling the entire pipeline from dataset creation to model design and evaluation. We assert that the future of ASQA lies not in more accurate single-score prediction, but in generating feedback that is as insightful and actionable as that of a human expert.

\section{Methodology}
Our framework is built on three pillars: (1) a novel, richly annotated dataset designed to capture the nuances of singing performance and uncover the inherent challenges of subjective assessment; (2) an exploration of model architectures, born from systematic experimentation, designed to overcome the practical limitations of analyzing full-length songs; and (3) a new human-in-the-loop benchmark for more meaningful and robust evaluation.

\subsection{The Sing-MD Dataset and Problem Definition}
To move beyond simple scoring, a dataset with rich, multi-faceted annotations is required. We constructed the Singing Multi-Dimension (Sing-MD) dataset. Its construction followed a two-stage process, beginning with a broad exploratory phase of amateur annotation, the conclusions of which guided the subsequent, more focused professional annotation phase. This process also revealed a fundamental challenge in the field: the label-input mismatch problem.

\subsubsection{Data Collection and Pre-screening}

The raw data comprises over 100,000 full-length a cappella singing recordings, covering multiple popular Chinese pop songs. All data are of full-song length. The data originates from anonymous recordings by users of offline KTV (Karaoke) venue applications in several major Chinese cities during the first half of 2024 (January to June). Prior to data collection, all users were informed that their anonymous data might be used for research purposes and were given the option to decline.

From an initial pool of over 100,000 recordings, we first used automated scripts to filter out segments with extremely low a cappella volume. We then performed manual screening to exclude recordings with severe audio quality issues (e.g., very poor singing proficiency, popping, clipping, excessively short duration, non-solo singing, or a mix of singing and speaking), resulting in a dataset of 10,000 clips. This dataset includes 6,347 female vocal clips and 3,653 male vocal clips. Culturally, all users are from mainland China, and a manual check of age revealed coverage of adolescents, middle-aged, and elderly individuals.

To ensure that subsequent manual annotations could focus on higher-level artistic qualities such as timbre, rather than basic technical errors, we established a rigorous pre-screening phase. Preliminary experiments confirmed that a performer's mistakes in fundamental singing skills, like pitch and rhythm, significantly impact a listener's perception of timbre's aesthetic quality. Therefore, we employed a proprietary, rule-based singing scoring system, RuleSignal system (as shown in Figure~\ref{fig:rulesignal_interface}, explained below), which automates the scoring of basic skills like pitch and rhythm accuracy by comparing user recordings to the original studio versions. From the dataset of 10,000 recording clips, we used this system to select the top 10\% of performances by score, ultimately obtaining approximately 1,000 recordings that demonstrated high technical proficiency. This pre-screening process ensures that the data proceeding to the manual annotation stage is already of a high technical caliber, thereby allowing annotators to concentrate on evaluating more subjective and abstract artistic dimensions like timbre and emotional expression.

\subsubsection{A Brief Introduction to the RuleSignal Reference-based Singing Assessment System} \label{sec:rulesignal}
\begin{figure}[h]
  \centering
  \includegraphics[width=0.99\columnwidth]{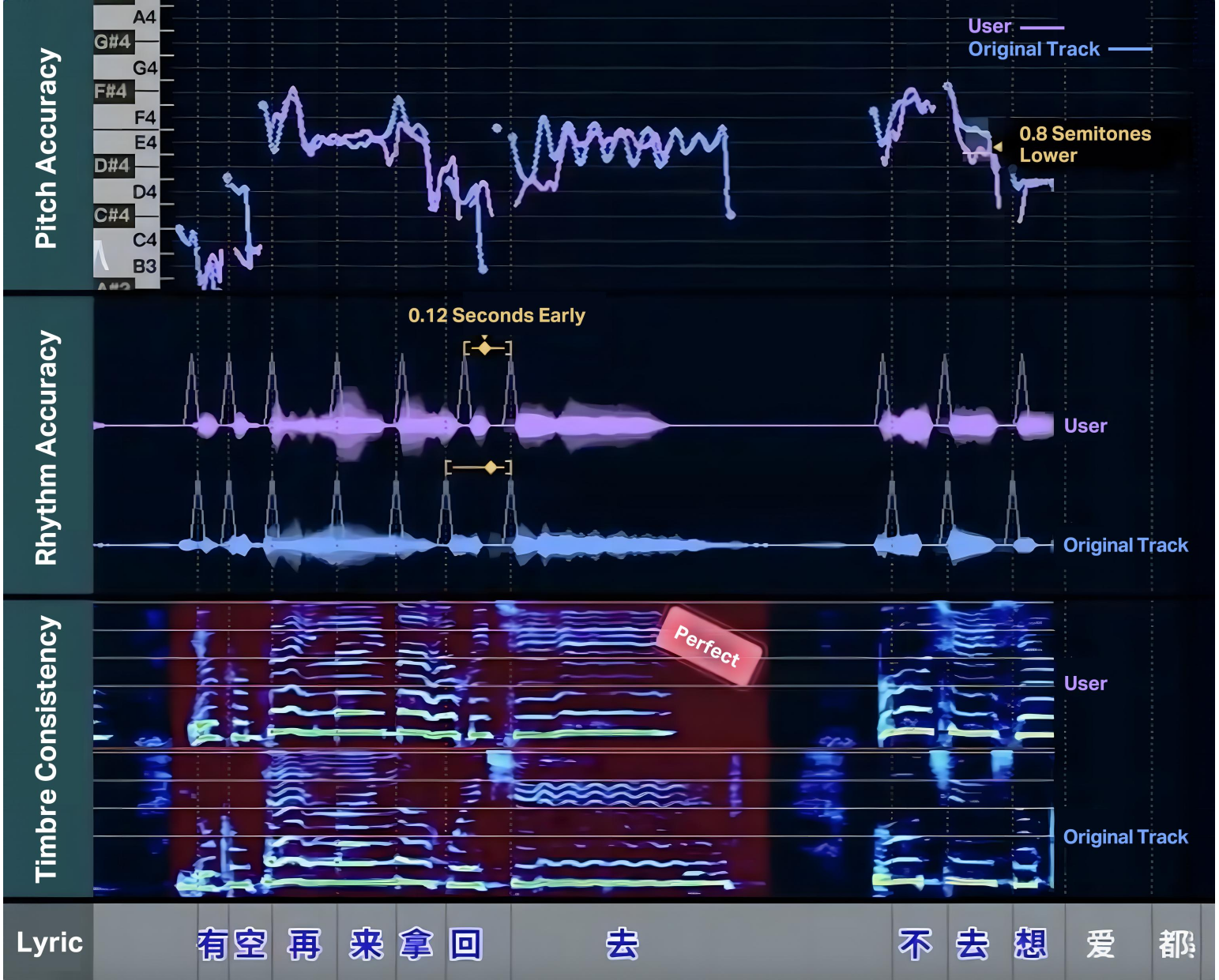}
  \Description{The image displays the user interface of the RuleSignal system, which is divided into three horizontal panels for evaluating a singing performance. The top panel, labeled 'Pitch Accuracy,' shows the user's pitch contour (in purple) overlaid on the original track's pitch contour (in blue), against a musical staff background. An annotation indicates a deviation, for instance, '0.8 Semitones Lower'. The middle panel, 'Rhythm Accuracy,' compares the user's audio waveform (purple) with the original's (blue), highlighting timing differences with labels like '0.12 Seconds Early'. The bottom panel, 'Timbre Consistency,' shows the spectrogram of the user's voice above the original's, with a badge like 'Perfect' indicating high similarity. Below all panels, a lyric bar displays the corresponding Chinese lyrics.}
  \caption{A schematic of the RuleSignal system interface used for data pre-screening. It selects technically proficient recordings by comparing the user's performance (purple) against the original track (blue) on pitch, rhythm, and timbre consistency.}
  \label{fig:rulesignal_interface}
\end{figure}

RuleSignal is a rule-based, reference-dependent singing evaluation system used for screening recordings. Its core principle is to compare a user's singing (visualized in purple on the interface) with the original song vocal track (visualized in blue) across key acoustic dimensions. A UI demonstration video of the system is available\footnote{\url{https://drive.google.com/file/d/16y9ydu7PV2HTXtM4P9qjKIvS_FfCRN5k/view?usp=share_link}}. A schematic of the RuleSignal interface is shown in Figure~\ref{fig:rulesignal_interface}. 

The system primarily evaluates the following aspects:
\begin{itemize}[leftmargin=*]
    \item \textbf{Pitch Accuracy}: The system extracts the pitch contours (F0 curves) of both the user's and the original singing. After aligning the two curves using the Dynamic Time Warping (DTW) algorithm, it calculates the difference between them. The system can automatically handle systematic differences in the overall pitch range between the user and the original, focusing on the accuracy of relative pitch.
    \item \textbf{Rhythm Accuracy}: The system analyzes the rhythmic consistency of the user's singing with the original by detecting note onsets and offsets. It compares whether the start and end times of each note or syllable in the user's performance match those in the corresponding positions of the original track.
    \item \textbf{Timbre Consistency (preliminary)}: During the screening phase, to exclude samples with styles that deviate too much from the original, the system also performs a preliminary comparison of the similarity in features like Mel-spectrograms between the user and the original to assess the general stylistic consistency of timbre features.
\end{itemize}
By integrating scores from these dimensions, RuleSignal can filter out recordings that meet a certain standard in basic singing skills (especially pitch and rhythm), thus providing a pool of relatively high-quality candidates for the subsequent, more detailed manual timbre evaluation, allowing the annotation to more closely approximate an assessment of \textbf{``pure timbre''}.

% The principle of pitch and rhythm evaluation is illustrated in Figure~\ref{fig:rulesignal_pitch}.

% \begin{figure}[h]
%   \centering
%   \includegraphics[width=0.99\columnwidth]{PIC/rulesignal-pitch.pdf}
%   \caption{Principle diagram of reference-based singing evaluation in the RuleSignal system. After aligning the two curves (original in red, user in blue) using the Dynamic Time Warping (DTW) algorithm, it calculates the difference. The system compares the pitch and rhythm relationship, automatically ignoring overall octave differences.}
%   \label{fig:rulesignal_pitch}
% \end{figure}

\subsubsection{Vocal Separation and Enhancement}
Subsequently, each selected recording underwent vocal separation and enhancement. We employed an advanced deep learning-based source separation model to extract clean, dry vocals from user recordings that might contain light accompaniment or reverberation.

The specific processing workflow is as follows:
\begin{enumerate}
    \item \textbf{Vocal and Accompaniment Separation:} First, the state-of-the-art music source separation model MelBand Roformer (viperx edition)\footnote{Model based on \url{https://github.com/ZFTurbo/Music-Source-Separation-Training}} is used to separate vocals and remove background noise.
    \item \textbf{De-reverberation:} Next, the vocals separated in the previous step are processed using the MelBand Roformer (anvuew edition) dereverb model to eliminate any residual reverberation effects.
\end{enumerate}

\textbf{Output:} The final result is clean, monophonic, pure vocal waveform data.
Before proceeding to multi-dimensional annotation, all audio was once again manually verified to ensure it was solo a cappella singing with no significant background noise or interference from the original track.

\subsubsection{Preliminary Exploration with Amateur Annotation}
We initially attempted to capture popular aesthetic preferences by organizing a large-scale annotation effort with 165 non-music-major annotators. Aged between 20 and 45, all were singing enthusiasts with broad exposure to and good auditory discrimination skills for pop music, representing the preferences of the wider Chinese audience regarding timbre popularity. All annotators participated in a two-day calibration training before formal annotation, where they aligned their scoring standards by trial-annotating 10 identical songs to ensure a consistent understanding of the criteria. Each audio clip was randomly assigned to annotators for independent, anonymous rating. Annotators were required to listen in a quiet environment using high-fidelity headphones. The scoring criterion was ``degree of timbre popularity'' on a 1-5 Likert scale (1 for very unpopular, 5 for very popular). They were explicitly instructed to focus on the inherent qualities of the voice itself (e.g., roundness, brightness, texture) and to minimize the influence of non-timbre factors such as the performer's pitch accuracy, rhythm, lyrical pronunciation, or familiarity with the song.

\paragraph{Annotation Design to Mitigate Bias.} The 1,000 pre-screened recordings included 33 songs, with each song having approximately 30-35 different user-sung versions. We recognized that an annotator's personal preference for a specific song could heavily bias their rating of the performance itself. To mitigate this bias, we implemented a strict annotation protocol: each annotator could only rate all user performances for a single song, with cross-song annotation being prohibited. This design (33 songs $\times$ 5 annotators/song = 165 annotators) ensures that the ratings reflect the quality of the performance rather than the annotator's fondness for the musical work itself.

\paragraph{Forced Distribution and Its Findings.} Each recording was rated by 5 amateur annotators on ``overall pleasantness'' (1-5 scale). To prevent score clustering, annotators were required to use a forced distribution method, ensuring that each score from 1 to 5 was used roughly evenly within the batch of performances they annotated. A check by quality inspectors confirmed that the vast majority of annotators complied with this rule, distributing their scores in a 1:1:1:1:1 ratio.

However, after aggregating the scores (by averaging the 5 ratings), we discovered a critical issue: the vast majority of the final average scores clustered in the middle range of 2-4. Due to diverse aesthetic tastes, an extreme score (1 or 5) given by one annotator was often neutralized by the scores from others. This led to an imbalanced score distribution in the final dataset, lacking sufficient samples at the quality extremes. This finding indicated that while amateur crowdsourcing is useful, the inherent variance in aesthetic preference makes it unreliable for creating a well-distributed dataset for nuanced, multi-dimensional training. This prompted us to turn to expert annotators.

\subsubsection{Professional Multi-Dimensional Annotation}
Based on the findings from the amateur annotation phase, we initiated a more focused annotation task conducted by two professional vocal coaches. Their professional backgrounds include being senior vocal teachers and music producers with extensive experience in vocal evaluation.

\paragraph{Annotation Schema.} We designed the annotation dimensions based on vocal pedagogy, with detailed criteria shown in Table~\ref{tab:scoring_criteria}. Among these, vocal technique and emotional expression are generally considered to be significantly improvable through training, whereas the timbre score, though also potentially influenced by vocal training, is largely related to an individual's physiological and acoustic characteristics and exhibits relative stability~\cite{sundberg1974articulatory}.
In addition to a 1-5 integer score, experts were required to provide a textual description, writing a short critique for each dimension. This dual-modality annotation (score + text) is crucial for training our descriptive models.

\begin{table}[h]
  \caption{Scoring criteria provided to expert annotators for the four dimensions.}
  \label{tab:scoring_criteria}
  \centering
  \resizebox{\columnwidth}{!}{%
  \begin{tabular}{l|l}
    \toprule
    \textbf{Dimension} & \textbf{Scoring Criteria (1=Poor, 5=Excellent)} \\
    \midrule
    \textbf{Vocal Technique} & 1: Obvious errors in pitch/rhythm. \\
    & 3: Technically correct but lacks highlights. \\
    & 5: Skillful, with masterful use of techniques. \\
    \hline
    \textbf{Emotional Expression} & 1: Mechanical, emotionally detached. \\
    & 3: Basic emotion conveyed, but lacks depth. \\
    & 5: Highly expressive, strong resonance with listener. \\
    \hline
    \textbf{Timbre Quality} & 1: Tone is flat or conflicts with the song's style. \\
    & 3: Clear and recognizable, but lacks variation. \\
    & 5: Unique tone with rich texture and layers. \\
    \hline
    \textbf{Breath Control} & 1: Unstable, with audible or disruptive breaths. \\
    & 3: Generally stable, but lacks fine control. \\
    & 5: Effortless control, supporting complex phrases. \\
    \bottomrule
  \end{tabular}%
  }
\end{table}

\subsubsection{Key Finding: Inter-Annotator Disagreement}
To quantify the subjectivity of the task, we had two experts annotate the same set of 30 performances. The results, shown in Table~\ref{tab:annotator_disagreement}, revealed that the exact agreement rate was consistently below 45\%. In a preliminary experiment, a model trained on mixed data (A+B) showed a smaller difference from expert A's scores than the difference between expert B and expert A. This strongly demonstrates that a single ``ground-truth'' score is a statistical fiction, invalidating the use of MAE or accuracy as primary evaluation metrics and motivating our development of a new evaluation paradigm.

\begin{table}[h]
  \caption{Inter-Annotator Agreement between two experts on the same 30 performances. ``Exact'' refers to the percentage of times scores were identical. ``$\pm$1'' refers to the percentage of times scores were within one point of each other.}
  \label{tab:annotator_disagreement}
  \centering
  \begin{tabular}{lcc}
    \toprule
    Dimension & Agreement (\%) & $\pm$1 Agreement (\%) \\
    \midrule
    Breath Control & 43.7 & 80.0 \\
    Timbre Quality & 37.5 & 70.0 \\
    Emotional Expression & 28.1 & 83.0 \\
    Vocal Technique & 35.2 & 90.0 \\
    \bottomrule
  \end{tabular}
\end{table}

\subsection{Model Architecture Exploration}
To address the challenge of full-song analysis, our methodology involves a systematic exploration of architectures, transitioning from clip-based MLLM approaches to more efficient, encoder-only models suitable for complete audio.

\subsubsection{The Label-Input Mismatch Problem: From Short Clips to Full Songs}
A major technical hurdle in analyzing singing is the mismatch between the scope of annotation and the model's input. Experts provide a score based on the \textit{entire} song (3-5 minutes), but standard MLLMs like Qwen-Audio~\cite{Qwen2Audio} can only process short clips (e.g., 30 seconds) due to prohibitive GPU memory requirements on consumer-grade hardware (e.g., a single RTX 4090). This creates a critical \textbf{label-input mismatch} problem: a song might receive a low score due to a flawed chorus, but the model may only be trained on a perfectly fine verse, leading to a confusing training signal. This mismatch prevents the model from learning the true relationship between vocal details and quality, causing it to learn spurious correlations and fail to generalize. Therefore, our core technical challenge was to develop a model capable of processing full-length songs.

\subsubsection{Clip-based MLLM Baselines}
Despite recognizing the label-input mismatch problem with short clips, our first step was to establish a baseline for standard MLLM fine-tuning methods. We explored three variants of fine-tuning Qwen2-Audio-7B on 30-second audio clips to evaluate the capability of MLLM architectures under this constraint:
\begin{itemize}[leftmargin=*]
    \item \textbf{A1: Direct Generation.} Fine-tuning the model using LoRA (Low-Rank Adaptation)~\cite{lora} to generate score tokens ('1'-'5'), with the final score determined by the probability distribution of these tokens in the output.
    \item \textbf{A2: LoRA + MLP.} Fine-tuning with LoRA, then feeding the features from the LLM's final output layer into a trainable MLP head for score classification.
    \item \textbf{A3: Frozen LLM + MLP.} Without fine-tuning, directly extracting the hidden state embeddings from the final layer of the LLM to train a separate MLP classifier.
\end{itemize}
All three methods employed fixed prompt engineering during training. We designed suitable text prompts to guide the model to generate the desired output.
As will be shown in the experiments section, all three clip-based methods performed poorly and yielded unstable results, confirming the necessity of a full-song analysis approach.

\subsubsection{Full-Song Scoring Architecture (Encoder-Only)}

\paragraph{Audio Foundation Model for Feature Extraction.}
To enable final deployment on consumer-grade 4090 GPUs for inference on full songs up to 3 minutes long without causing memory overflow, we abandoned the LLM component of Qwen2-Audio-7B for scoring and focused on two more efficient audio encoders.

\begin{enumerate}[leftmargin=*]
    \item \textbf{Encoder-Qwen:} The standalone audio encoder from the Qwen2-Audio model, based on Whisper-v3~\cite{Whisper}.
    \item \textbf{Encoder-SaMoye:} An enhanced Whisper encoder from our previous work on singing voice conversion (SaMoyeSVC~\cite{wang2024samoyearxiv}). This encoder was pre-trained on 1700 hours of singing data and incorporates additional features like PPG, contentvec, and F0 embeddings, potentially providing richer representations for singing.
\end{enumerate}

These encoders are responsible for extracting high-quality, infor-mation-rich, general acoustic representations from the pre-processed clean vocals. A key advantage is that these encoders can be fully trained end-to-end on an 8-card 4090 GPU cluster, without requiring LoRA like the LLM component of Qwen2-Audio-7B. Furthermore, they can process long audio sequences by taking complete clean vocal waveform segments (in 2-5 second windows with a certain stride) as input while consuming relatively little VRAM.

\paragraph{Downstream Scoring Modules.}
We experimented with four different downstream modules following the full-song encoder to produce the final score:
\begin{enumerate}[leftmargin=*]
    \item \textbf{MLP Classifier:} A simple baseline that applies average pooling over time to the sequence of encoder feature embeddings and passes them to a classification head composed of fully connected layers (MLP).
    \item \textbf{RNN-based Classifier:} Uses an RNN (e.g., LSTM/GRU) to model the temporal dependencies in the feature sequence before classification.
    \item \textbf{Transformer-based Classifier:} Uses a Transformer encoder to capture long-range dependencies in the feature sequence, followed by pooling and classification.
\end{enumerate}
    
All these scoring modules are trained using the manual annotations as the supervisory signal.

\subsection{Descriptive Feedback Generation}
Generating textual feedback faces the same 30-second limitation as clip-based scoring models. To address this, we adopted a pragmatic multi-stage strategy for training and inference.

\paragraph{Training.} We fine-tuned the Qwen2-Audio-7B model on the Sing-MD dataset using LoRA. Training instances consisted of random 30-second audio clips paired with their corresponding expert textual critiques. While this does not fully resolve the full-song label-input mismatch, it enables the model to learn the association between local audio events and descriptive language.

\paragraph{Inference.} For a full-length song, we segment it into multiple 30-second clips. The LoRA-finetuned model generates a diagnostic critique for each clip. These individual critiques are then concatenated to form a comprehensive list of identified issues spanning the entire performance. This aggregated text serves as a rich diagnostic input for the final guidance and suggestion stage in our system pipeline, which is polished and summarized by a much larger LLM API.

\section{Experiments}
This section details the experimental validation of our designed framework. Our experimental design aims to answer the following three key research questions, which follow the narrative logic of our methodological development:
\begin{enumerate}[leftmargin=*,nosep]
    \item How severe is the label-input mismatch problem? Can clip-based MLLM baselines effectively handle this task?
    \item Within our developed full-song architecture, which combination of encoder and downstream module performs best across the four professional dimensions?
    \item How does the best-performing model fare on our designed H-TPR benchmark? Does this evaluation align with human perceptual judgment?
\end{enumerate}

\subsection{Experimental Setup}

\subsubsection{Dataset and Splits}
All experiments were conducted using the \textbf{professional subset} of our created \textbf{Sing-MD} dataset. This subset contains 1,000 recordings with four-dimensional scores and textual critiques from a single expert. We used an 80/10/10 split for the training, validation, and test sets, ensuring no overlap at the user identity level to prevent data leakage.

\subsubsection{Baseline Models}
We compare our full-song architecture against several baseline models that represent different problem-solving approaches:

\begin{itemize}[leftmargin=*]
    \item \textbf{Clip-based MLLM Baselines:} The three variants described previously (Direct Generation, LoRA+MLP, Frozen+MLP), all fine-tuned on 30-second clips from our training data.
    \item \textbf{Adapted Standard ASQA Baselines:} We adapted several state-of-the-art ASQA models~\cite{Ju2024EndToEnd, Sun2023TGCritic, Chen2025TrainingBetterEmbedding} to fit our multi-dimensional scoring task. By modifying their output layers and retraining them on our dataset, we can more fairly compare the performance of these SOTA ideas on this task. The specific adaptations are as follows:
    \begin{itemize}[leftmargin=*]
        \item \textbf{Ju2024-Adapted}~\cite{Ju2024EndToEnd}: The original model is for overall singing skill assessment. To adapt it, we retained its CRNN+Attention structure and BiCA feature fusion mechanism, but changed the input features to CQT and X-vector, modified the output layer to a five-class classification head, and did not use its accompaniment mixing strategy.
        
        \item \textbf{Sun2023-Adapted}~\cite{Sun2023TGCritic}: The original model (TG-Critic) is a timbre-guided three-class singing assessment model. For adaptation, we kept its dual-branch structure but replaced the timbre embedding with a WeSpeaker embedding, changed the Softmax layer to five-class classification, and did not directly use its iterative self-labeling method.
        
        \item \textbf{Chen2025-Adapted}~\cite{Chen2025TrainingBetterEmbedding}: The original model learns a robust SQA embedding via perturbation-based data augmentation. For adaptation, we adopted its core CRNN structure and pitch histogram input, but modified the output layer to a five-class classification head and did not use its perturbation data augmentation or auxiliary classifier mechanisms.
    \end{itemize}

    \item \textbf{WeSpeaker+MLP:} A simple baseline using a pre-trained WeSpeaker to extract speaker embeddings~\cite{WeSpeaker} connected to an MLP classifier, used to test the effectiveness of relying solely on timbre features.
\end{itemize}

\subsubsection{Implementation Details}
All models were implemented in PyTorch and trained on an 8-card NVIDIA RTX 4090 GPU cluster. For MLLM fine-tuning, we used LoRA with a rank of 16. For encoder-only models, we used the AdamW optimizer with an initial learning rate of $1 \times 10^{-4}$ to fully fine-tune the encoders. All audio inputs were resampled to 16kHz.

\subsubsection{Evaluation Metric: From Score Matching to Perceptual Ranking}
As revealed in Section 3.1.4, there is significant disagreement among experts in scoring. This core finding indicates that traditional accuracy metrics—whether \textbf{Exact Accuracy} or \textbf{$\pm$1 Accuracy}—are misleading because they wrongly assume the existence of a stable ``ground-truth'' score.

Therefore, we designed a new evaluation paradigm where the goal shifts from ``How close is the model's predicted score to the expert's score?'' to a more fundamental and application-oriented question: ``Is the model's quality ranking of songs perceptually consistent and reasonable to a human listener?''

To realize this paradigm, we designed and adopted the **Human-in-the-loop Tiered Perceptual Ranking (H-TPR)** as our primary and sole evaluation benchmark. The H-TPR protocol is as follows:
\begin{enumerate}[leftmargin=*,nosep]
    \item \textbf{Tiering:} Based on the model's predicted scores for all songs in the test set, divide the songs into three quality tiers: High, Medium, and Low.
    \item \textbf{Sampling:} Randomly draw one song from each of the three tiers to form a (High, Medium, Low) quality triplet.
    \item \textbf{Blind Judgment:} A human evaluator listens to the triplet without knowledge of the scores or their origin and judges whether the three songs exhibit a clearly discernible quality gradient (i.e., ``High'' is indeed better than ``Medium,'' and ``Medium'' is indeed better than ``Low'').
\end{enumerate}
The final H-TPR score, which is the percentage of triplets judged by the evaluator as ``rank-consistent,'' directly quantifies the perceptual validity of the model's ranking results, which is the core requirement for real-world applications.

\subsection{Model Architecture Comparison and Optimal Selection}
Our core experiment aims to identify the most effective full-song architecture. On the professional subset of Sing-MD, we systematically evaluated combinations of two encoders (Encoder-Qwen, Encoder-SaMoye) and the first three downstream modules (MLP, RNN, Transformer). The performance of each combination was measured using our H-TPR benchmark.

Table~\ref{tab:main_results_htpr} shows the H-TPR scores for the best-performing encoder-module combination for each dimension. The results reveal a key finding: no single architecture is optimal across all dimensions. Specifically, the combination of Encoder-SaMoye with an RNN performed best on the two more technical dimensions of breath control and vocal technique, likely benefiting from its rich, singing-specific pre-training knowledge and the RNN's ability to model local temporal dynamics. In contrast, the more general Encoder-Qwen paired with a simple MLP classifier excelled on the more aesthetically-oriented dimensions of timbre quality and emotional expression, suggesting these dimensions may rely more on global, generalized acoustic features. Based on these results, we constructed the hybrid system VocalVerse, which integrates the best-performing model for each dimension.

\begin{table*}[h]
\caption{H-TPR scores (\%) of different full-song architecture combinations across the four professional dimensions. The results show that no single architecture is universally superior. The highest score for each dimension is marked in bold, and these best-performing models together form our final hybrid system, VocalVerse.}
\label{tab:main_results_htpr}
\centering
\begin{tabular}{lcccc}
\toprule
\textbf{Model Architecture} & \textbf{Breath Control} & \textbf{Timbre Quality} & \textbf{Emotional Expression} & \textbf{Vocal Technique} \\
\midrule
Encoder-Qwen + MLP           & 76.1            & \textbf{79.5} & \textbf{76.7} & 77.5             \\
Encoder-Qwen + RNN           & 77.5            & 77.4          & 75.5          & 79.7             \\
Encoder-Qwen + Transformer   & 74.9            & 75.7          & 73.5          & 75.9             \\
\hline
Encoder-SaMoye + MLP         & 80.1            & 78.5          & 76.3          & 81.4             \\
Encoder-SaMoye + RNN         & \textbf{82.4}   & 78.2          & 74.8          & \textbf{82.9}    \\
Encoder-SaMoye + Transformer & 80.8            & 77.0          & 73.8          & 80.6             \\
\bottomrule
\end{tabular}
\end{table*}

% \begin{table*}[t]
% \caption{H-TPR scores (\%) of different full-song architecture combinations across the four professional dimensions. The results show that no single architecture is universally superior. The highest score for each dimension is marked in bold, and these best-performing models together form our final hybrid system, VocalVerse.}
% \label{tab:main_results_htpr}
% \centering
% \begin{tabularx}{\textwidth}{l | c c c c}
% \toprule
% \textbf{Model Architecture} & \textbf{Breath Control} & \textbf{Timbre Quality} & \textbf{Emotional Expression} & \textbf{Vocal Technique} \\
% \midrule
% Encoder-Qwen + MLP & 76.1 & \textbf{79.5} & \textbf{76.7} & 77.5 \\
% Encoder-Qwen + RNN & 77.5 & 77.4 & 75.5 & 79.7 \\
% Encoder-Qwen + Transformer & 74.9 & 75.7 & 73.5 & 75.9 \\
% \midrule
% Encoder-SaMoye + MLP & 80.1 & 78.5 & 76.3 & 81.4 \\
% Encoder-SaMoye + RNN & \textbf{82.4} & 78.2 & 74.8 & \textbf{82.9} \\
% Encoder-SaMoye + Transformer & 80.8 & 77.0 & 73.8 & 80.6 \\
% \bottomrule
% \end{tabularx}
% \end{table*}

\subsection{Main Results: Comparison with Baselines}
To validate the effectiveness of our final VocalVerse system, we conducted a comprehensive H-TPR performance comparison against a series of representative baseline models across all four professional dimensions. These baselines include: (1) mainstream ASQA architectures; (2) a simple model relying only on timbre features; and (3) clip-based MLLM models used to verify the necessity of full-song analysis.

As shown in Table~\ref{tab:final_comparison}, VocalVerse significantly outperforms all baseline models across all evaluation dimensions. This result strongly demonstrates the effectiveness of our full-song analysis framework and the correctness of our hybrid strategy of using the optimal model for each different dimension.

Among all baselines, the clip-based MLLM models performed particularly poorly, providing the most direct quantitative evidence for the necessity of full-song analysis. We observed that their core issue lies in a lack of stable generalization: a model trained on one set of random 30-second clips performed poorly on another set of random clips from the same test songs. This clearly confirms how the ``label-input mismatch'' problem hinders the model from learning true vocal quality relationships, forcing it to rely on spurious correlations, which ultimately has a fatal impact on model performance.

\begin{table*}[ht]
\caption{Comparison of H-TPR scores (\%) between our VocalVerse system and various baseline models across all four dimensions.}
\label{tab:final_comparison}
\centering
\begin{tabular}{lcccc}
\toprule
\textbf{Model} & \textbf{Breath Control} & \textbf{Timbre Quality} & \textbf{Emotional Expression} & \textbf{Vocal Technique} \\
\midrule
\multicolumn{5}{l}{\textit{Standard Baseline Models}} \\
WeSpeaker+MLP & 58.9 & 63.1 & 61.5 & 56.6 \\
Ju2024-Adapted & 70.4 & 68.0 & 65.8 & 73.4 \\
Sun2023-Adapted & 68.3 & 71.8 & 68.8 & 70.2 \\
Chen2025-Adapted & 72.3 & 69.3 & 67.1 & 74.9 \\
\hline
\multicolumn{5}{l}{\textit{Clip-based MLLM Baselines}} \\
A1: Direct Generation & 47.8 & 50.2 & 52.2 & 46.5 \\
A2: LoRA + MLP & 54.5 & 56.1 & 57.7 & 53.4 \\
A3: Frozen LLM + MLP & 52.4 & 54.1 & 55.7 & 51.1 \\
\hline
\textbf{VocalVerse (ours)} & \textbf{82.4} & \textbf{79.5} & \textbf{76.7} & \textbf{82.9} \\
\bottomrule
\end{tabular}
\end{table*}

\section{Discussion}
Our work represents a fundamental shift in how automatic singing assessment is conceptualized and executed. By moving from predicting a single, noisy score to providing a multi-dimensional, descriptive evaluation, we align the task more closely with human expertise and practical user needs. Our key finding on expert disagreement underscores the futility of chasing perfect accuracy on a subjective task and validates our pivot to a ranking-based evaluation benchmark, H-TPR. The development of the VocalVerse architecture demonstrates that it is possible to perform complex, full-song analysis on consumer hardware by departing from the paradigm of monolithic MLLMs and embracing efficient, specialized modules.

Furthermore, our pre-screening methodology, which filters for basic technical proficiency, also serves to mitigate the influence of confounding factors like poor pitch and rhythm on the assessment of higher-level artistic qualities such as timbre. This addresses a known challenge in the field, ensuring that our models evaluate artistry rather than just correcting basic mistakes.

\subsection{Limitations}
Our work is not without limitations. Firstly, regarding the data, while the Sing-MD dataset is substantial, its origin from Chinese KTV  (Karaoke) recordings leads to a significant lack of diversity in areas such as language (currently only Chinese), musical genres, and singing styles. This not only limits the model's generalizability to broader cultural contexts but also implies that our findings on expert disagreement may be culturally specific. Secondly, concerning the evaluation method, while the H-TPR benchmark is more meaningful, its labor-intensive nature limits its scalability compared to automated metrics.

% 太长了没用\subsection{Future Work: From Scoring to Ranking}
% A key limitation of any scoring system, including ours, is the compression of quality into a single number per dimension. In high-stakes scenarios like competitions, this often leads to score inflation and a lack of differentiability at the top end. We believe the next frontier is to move from \textbf{scoring} to direct \textbf{ranking}. Our future work will explore a new paradigm based on pairwise comparison. By training a model on a large dataset of ``Which of these two performances is better?'' judgments, we can learn to directly output a ranked list. This approach is better suited for competitive applications and represents a more direct way of modeling relative human preference.

\subsection{Future Work: From Scoring to Ranking}
For future work, we propose shifting from scoring to direct ranking. By training a model on a large dataset of pairwise preference judgments (e.g., ``A is better than B''), we can learn to generate ranked lists directly. This approach is more suitable for competitive applications as it more directly models relative human preference.

\section{Conclusion}
In this paper, we challenged the prevailing paradigm of automatic singing assessment. We argued that the pursuit of a single, objective score is a flawed goal for an inherently subjective and multi-faceted skill. We introduced a comprehensive framework built on three pillars: the \textbf{Sing-MD} dataset, which revealed the fundamental challenge of expert disagreement; the \textbf{VocalVerse} model, an efficient hybrid architecture for full-song analysis; and the \textbf{H-TPR} benchmark, a human-centric method for evaluating a model's ability to create perceptually valid rankings. Our system not only outperforms existing methods on this new, more meaningful benchmark but has also proven its value through large-scale commercial deployment in over 800 karaoke venues, serving more than 200,000 users monthly. By shifting the focus from scoring to descriptive understanding, our work lays the groundwork for a new generation of AI-powered tools for music education and artistic development.
% , and points the way toward a future of direct, preference-based ranking.

\begin{acks}
This work was supported by the National Natural Science Foundation of China (No.62272409).
\end{acks}

\bibliographystyle{ACM-Reference-Format}
\balance
\bibliography{sample-base}


\begin{thebibliography}{42}

%%% the Web address.

\ifx \showCODEN    \undefined \def \showCODEN     #1{\unskip}     \fi
\ifx \showISBNx    \undefined \def \showISBNx     #1{\unskip}     \fi
\ifx \showISBNxiii \undefined \def \showISBNxiii  #1{\unskip}     \fi
\ifx \showISSN     \undefined \def \showISSN      #1{\unskip}     \fi
\ifx \showLCCN     \undefined \def \showLCCN      #1{\unskip}     \fi
\ifx \shownote     \undefined \def \shownote      #1{#1}          \fi
\ifx \showarticletitle \undefined \def \showarticletitle #1{#1}   \fi
\ifx \showURL      \undefined \def \showURL       {\relax}        \fi
% invisible to TeX
\providecommand\bibfield[2]{#2}
\providecommand\bibinfo[2]{#2}
\providecommand\natexlab[1]{#1}
\providecommand\showeprint[2][]{arXiv:#2}

\bibitem[Chen and Soo(2025)]%
        {Chen2025TrainingBetterEmbedding}
\bibfield{author}{\bibinfo{person}{Po-Wei Chen} {and} \bibinfo{person}{Von-Wun Soo}.} \bibinfo{year}{2025}\natexlab{}.
\newblock \showarticletitle{Training Better Embedding With Perturbed Data Augmentation for Automatic Singing Quality Assessment}. In \bibinfo{booktitle}{\emph{2025 IEEE International Conference on Acoustics, Speech and Signal Processing (ICASSP)}}. \bibinfo{pages}{1--5}.
\newblock
\href{https://doi.org/10.1109/ICASSP49660.2025.10888711}{doi:\nolinkurl{10.1109/ICASSP49660.2025.10888711}}
\newblock
\shownote{Forthcoming, URL: \url{https://doi.org/10.1109/ICASSP49660.2025.10888711}}.


\bibitem[Chu et~al\mbox{.}(2023)]%
        {Qwen2Audio}
\bibfield{author}{\bibinfo{person}{Yunfei Chu}, \bibinfo{person}{Jin Xu}, \bibinfo{person}{Xiaohuan Zhou}, \bibinfo{person}{Qian Yang}, \bibinfo{person}{Shiliang Zhang}, \bibinfo{person}{Zhijie Yan}, \bibinfo{person}{Chang Zhou}, {and} \bibinfo{person}{Jingren Zhou}.} \bibinfo{year}{2023}\natexlab{}.
\newblock \showarticletitle{Qwen-audio: Advancing universal audio understanding via unified large-scale audio-language models}.
\newblock \bibinfo{journal}{\emph{arXiv preprint arXiv:2311.07919}} (\bibinfo{year}{2023}).
\newblock


\bibitem[Copet et~al\mbox{.}(2024)]%
        {copet2024simple}
\bibfield{author}{\bibinfo{person}{Jade Copet}, \bibinfo{person}{Felix Kreuk}, \bibinfo{person}{Itai Gat}, \bibinfo{person}{Tal Remez}, \bibinfo{person}{David Kant}, \bibinfo{person}{Gabriel Synnaeve}, \bibinfo{person}{Yossi Adi}, {and} \bibinfo{person}{Alexandre D{\'e}fossez}.} \bibinfo{year}{2024}\natexlab{}.
\newblock \showarticletitle{Simple and controllable music generation}.
\newblock \bibinfo{journal}{\emph{Advances in Neural Information Processing Systems}}  \bibinfo{volume}{36} (\bibinfo{year}{2024}).
\newblock


\bibitem[Dai et~al\mbox{.}(2025)]%
        {dai2025everyone}
\bibfield{author}{\bibinfo{person}{Shuqi Dai}, \bibinfo{person}{Yunyun Wang}, \bibinfo{person}{Roger~B Dannenberg}, {and} \bibinfo{person}{Zeyu Jin}.} \bibinfo{year}{2025}\natexlab{}.
\newblock \showarticletitle{Everyone-Can-Sing: Zero-Shot Singing Voice Synthesis and Conversion with Speech Reference}. In \bibinfo{booktitle}{\emph{ICASSP 2025-2025 IEEE International Conference on Acoustics, Speech and Signal Processing (ICASSP)}}. IEEE, \bibinfo{pages}{1--5}.
\newblock


\bibitem[Dhariwal et~al\mbox{.}(2020)]%
        {dhariwal2020jukebox}
\bibfield{author}{\bibinfo{person}{Prafulla Dhariwal}, \bibinfo{person}{Heewoo Jun}, \bibinfo{person}{Christine Payne}, \bibinfo{person}{Jong~Wook Kim}, \bibinfo{person}{Alec Radford}, {and} \bibinfo{person}{Ilya Sutskever}.} \bibinfo{year}{2020}\natexlab{}.
\newblock \showarticletitle{Jukebox: A generative model for music}.
\newblock \bibinfo{journal}{\emph{arXiv preprint arXiv:2005.00341}} (\bibinfo{year}{2020}).
\newblock


\bibitem[Gupta et~al\mbox{.}(2020)]%
        {Gupta2020TwinNeural}
\bibfield{author}{\bibinfo{person}{Chitralekha Gupta}, \bibinfo{person}{Lin Huang}, {and} \bibinfo{person}{Haizhou Li}.} \bibinfo{year}{2020}\natexlab{}.
\newblock \showarticletitle{Automatic rank-ordering of singing vocals with twin-neural network}. In \bibinfo{booktitle}{\emph{Proceedings of the 21st International Society for Music Information Retrieval Conference (ISMIR)}}. \bibinfo{pages}{416--423}.
\newblock
\urldef\tempurl%
\url{https://archives.ismir.net/ismir2020/paper/000165.pdf}
\showURL{%
\tempurl}


\bibitem[Gupta et~al\mbox{.}(2018)]%
        {Gupta2018TechnicalFramework}
\bibfield{author}{\bibinfo{person}{Chitralekha Gupta}, \bibinfo{person}{Haizhou Li}, {and} \bibinfo{person}{Ye Wang}.} \bibinfo{year}{2018}\natexlab{}.
\newblock \showarticletitle{A technical framework for automatic perceptual evaluation of singing quality}.
\newblock \bibinfo{journal}{\emph{APSIPA Transactions on Signal and Information Processing}}  \bibinfo{volume}{7}, Article \bibinfo{articleno}{e10} (\bibinfo{year}{2018}).
\newblock
\href{https://doi.org/10.1017/atsip.2018.10}{doi:\nolinkurl{10.1017/atsip.2018.10}}


\bibitem[Gupta et~al\mbox{.}(2021)]%
        {Gupta2021Rhythm}
\bibfield{author}{\bibinfo{person}{Chitralekha Gupta}, \bibinfo{person}{Jinhu Li}, {and} \bibinfo{person}{Haizhou Li}.} \bibinfo{year}{2021}\natexlab{}.
\newblock \showarticletitle{Towards reference-independent rhythm assessment of solo singing}. In \bibinfo{booktitle}{\emph{2021 Asia-Pacific Signal and Information Processing Association Annual Summit and Conference (APSIPA ASC)}}. \bibinfo{pages}{912--919}.
\newblock
\href{https://doi.org/10.1109/APSIPAASC53192.2021.9687795}{doi:\nolinkurl{10.1109/APSIPAASC53192.2021.9687795}}


\bibitem[Hu et~al\mbox{.}(2022)]%
        {lora}
\bibfield{author}{\bibinfo{person}{Edward~J Hu}, \bibinfo{person}{Yelong Shen}, \bibinfo{person}{Phillip Wallis}, \bibinfo{person}{Zeyuan Allen-Zhu}, \bibinfo{person}{Yuanzhi Li}, \bibinfo{person}{Shean Wang}, \bibinfo{person}{Lu Wang}, \bibinfo{person}{Weizhu Chen}, {et~al\mbox{.}}} \bibinfo{year}{2022}\natexlab{}.
\newblock \showarticletitle{Lora: Low-rank adaptation of large language models.}
\newblock \bibinfo{journal}{\emph{ICLR}} \bibinfo{volume}{1}, \bibinfo{number}{2} (\bibinfo{year}{2022}), \bibinfo{pages}{3}.
\newblock


\bibitem[Huang et~al\mbox{.}(2020)]%
        {Huang2020SpectralFeatures}
\bibfield{author}{\bibinfo{person}{Lin Huang}, \bibinfo{person}{Chitralekha Gupta}, {and} \bibinfo{person}{Haizhou Li}.} \bibinfo{year}{2020}\natexlab{}.
\newblock \showarticletitle{Spectral features and pitch histogram for automatic singing quality evaluation with {CRNN}}. In \bibinfo{booktitle}{\emph{2020 Asia-Pacific Signal and Information Processing Association Annual Summit and Conference (APSIPA ASC)}}. \bibinfo{pages}{492--499}.
\newblock
\urldef\tempurl%
\url{http://www.apsipa.org/proceedings/2020/pdfs/0000492.pdf}
\showURL{%
\tempurl}


\bibitem[Ju et~al\mbox{.}(2023)]%
        {Ju2023ImprovingASSE}
\bibfield{author}{\bibinfo{person}{Yaolong Ju}, \bibinfo{person}{Chen Xu}, \bibinfo{person}{Yuxuan Guo}, \bibinfo{person}{Jinhu Li}, {and} \bibinfo{person}{Simon Lui}.} \bibinfo{year}{2023}\natexlab{}.
\newblock \showarticletitle{Improving Automatic Singing Skill Evaluation with Timbral Features, Attention, and Singing Voice Separation}. In \bibinfo{booktitle}{\emph{2023 IEEE International Conference on Multimedia and Expo (ICME)}}. \bibinfo{pages}{612--617}.
\newblock
\href{https://doi.org/10.1109/ICME55011.2023.00111}{doi:\nolinkurl{10.1109/ICME55011.2023.00111}}


\bibitem[Ju et~al\mbox{.}(2024)]%
        {Ju2024EndToEnd}
\bibfield{author}{\bibinfo{person}{Yaolong Ju}, \bibinfo{person}{Jing Yang}, \bibinfo{person}{Chun~Yat Wu}, \bibinfo{person}{Betty~Corti{\~n}as Lorenzo}, \bibinfo{person}{Jiajun Deng}, \bibinfo{person}{Fan Fan}, {and} \bibinfo{person}{Simon Lui}.} \bibinfo{year}{2024}\natexlab{}.
\newblock \showarticletitle{End-to-end automatic singing skill evaluation using cross-attention and data augmentation for solo singing and singing with accompaniment}. In \bibinfo{booktitle}{\emph{Proceedings of the 25th International Society for Music Information Retrieval Conference (ISMIR)}}. \bibinfo{address}{San Francisco, United States}.
\newblock
\newblock
\shownote{Forthcoming, URL: \url{https://www.researchgate.net/publication/389749806}}.


\bibitem[Lee and Nam(2019)]%
        {Lee2019CROSS}
\bibfield{author}{\bibinfo{person}{Kyungyun Lee} {and} \bibinfo{person}{Juhan Nam}.} \bibinfo{year}{2019}\natexlab{}.
\newblock \showarticletitle{Learning a Joint Embedding Space of Monophonic and Mixed Music Signals for Singing Voice}. In \bibinfo{booktitle}{\emph{Proceedings of the 20th International Society for Music Information Retrieval Conference (ISMIR)}}. \bibinfo{address}{Delft, The Netherlands}, \bibinfo{pages}{295--302}.
\newblock


\bibitem[Li et~al\mbox{.}(2022)]%
        {Li2022VocEmb4SVS}
\bibfield{author}{\bibinfo{person}{Chenhao Li}, \bibinfo{person}{Yi Li}, \bibinfo{person}{Xiang Du}, \bibinfo{person}{Yaolong Ju}, \bibinfo{person}{Sihan Hu}, {and} \bibinfo{person}{Zhiyong Wu}.} \bibinfo{year}{2022}\natexlab{}.
\newblock \showarticletitle{{VocEmb4SVS}: Improving Singing Voice Separation with Vocal Embeddings}. In \bibinfo{booktitle}{\emph{2022 Asia-Pacific Signal and Information Processing Association Annual Summit and Conference (APSIPA ASC)}}. \bibinfo{pages}{234--239}.
\newblock
\href{https://doi.org/10.1109/APSIPAASC55918.2022.9980245}{doi:\nolinkurl{10.1109/APSIPAASC55918.2022.9980245}}


\bibitem[Li et~al\mbox{.}(2021)]%
        {Li2021TrainingExplainable}
\bibfield{author}{\bibinfo{person}{Jinhu Li}, \bibinfo{person}{Chitralekha Gupta}, {and} \bibinfo{person}{Haizhou Li}.} \bibinfo{year}{2021}\natexlab{}.
\newblock \showarticletitle{Training explainable singing quality assessment network with augmented data}. In \bibinfo{booktitle}{\emph{2021 Asia-Pacific Signal and Information Processing Association Annual Summit and Conference (APSIPA ASC)}}. \bibinfo{pages}{904--911}.
\newblock
\href{https://doi.org/10.1109/APSIPAASC53192.2021.9687793}{doi:\nolinkurl{10.1109/APSIPAASC53192.2021.9687793}}


\bibitem[Lian et~al\mbox{.}(2025)]%
        {lian2025affectgpt}
\bibfield{author}{\bibinfo{person}{Zheng Lian}, \bibinfo{person}{Haoyu Chen}, \bibinfo{person}{Lan Chen}, \bibinfo{person}{Haiyang Sun}, \bibinfo{person}{Licai Sun}, \bibinfo{person}{Yong Ren}, \bibinfo{person}{Zebang Cheng}, \bibinfo{person}{Bin Liu}, \bibinfo{person}{Rui Liu}, \bibinfo{person}{Xiaojiang Peng}, \bibinfo{person}{Jiangyan Yi}, {and} \bibinfo{person}{Jianhua Tao}.} \bibinfo{year}{2025}\natexlab{}.
\newblock \bibinfo{title}{AffectGPT: A New Dataset, Model, and Benchmark for Emotion Understanding with Multimodal Large Language Models}.
\newblock
\showeprint[arxiv]{2501.16566}~[cs.HC]
\urldef\tempurl%
\url{https://arxiv.org/abs/2501.16566}
\showURL{%
\tempurl}


\bibitem[Mao et~al\mbox{.}(2023)]%
        {10222365}
\bibfield{author}{\bibinfo{person}{Weihang Mao}, \bibinfo{person}{Bo Han}, {and} \bibinfo{person}{Zihao Wang}.} \bibinfo{year}{2023}\natexlab{}.
\newblock \showarticletitle{Sketchffusion: Sketch-Guided Image Editing with Diffusion Model}. In \bibinfo{booktitle}{\emph{2023 IEEE International Conference on Image Processing (ICIP)}}. \bibinfo{pages}{790--794}.
\newblock
\href{https://doi.org/10.1109/ICIP49359.2023.10222365}{doi:\nolinkurl{10.1109/ICIP49359.2023.10222365}}


\bibitem[McAdams(2009)]%
        {McAdams2009PerceptionTimbre}
\bibfield{author}{\bibinfo{person}{Stephen McAdams}.} \bibinfo{year}{2009}\natexlab{}.
\newblock \showarticletitle{The perception of musical timbre}.
\newblock In \bibinfo{booktitle}{\emph{The Oxford Handbook of Music Psychology}}, \bibfield{editor}{\bibinfo{person}{Susan Hallam}, \bibinfo{person}{Ian Cross}, {and} \bibinfo{person}{Michael Thaut}} (Eds.). \bibinfo{publisher}{Oxford University Press}, \bibinfo{pages}{72--80}.
\newblock


\bibitem[Nakano et~al\mbox{.}(2006a)]%
        {Nakano2006AutomaticEvaluation}
\bibfield{author}{\bibinfo{person}{Tomoyasu Nakano}, \bibinfo{person}{Masataka Goto}, {and} \bibinfo{person}{Yuzuru Hiraga}.} \bibinfo{year}{2006}\natexlab{a}.
\newblock \showarticletitle{An automatic singing skill evaluation method for unknown melodies using pitch interval accuracy and vibrato features}. In \bibinfo{booktitle}{\emph{Ninth International Conference on Spoken Language Processing (INTERSPEECH 2006)}}. \bibinfo{pages}{1706--1709}.
\newblock
\urldef\tempurl%
\url{https://www.isca-speech.org/archive/interspeech_2006/nakano06_interspeech.html}
\showURL{%
\tempurl}


\bibitem[Nakano et~al\mbox{.}(2006b)]%
        {Nakano2006Subjective}
\bibfield{author}{\bibinfo{person}{Tomoyasu Nakano}, \bibinfo{person}{Masataka Goto}, {and} \bibinfo{person}{Yuzuru Hiraga}.} \bibinfo{year}{2006}\natexlab{b}.
\newblock \showarticletitle{Subjective evaluation of common singing skills using the rank ordering method}. In \bibinfo{booktitle}{\emph{Proceedings of the 9th International Conference on Music Perception and Cognition (ICMPC9)}}, \bibfield{editor}{\bibinfo{person}{Mario Baroni}, \bibinfo{person}{Anna~Rita Addessi}, \bibinfo{person}{Roberto Caterina}, {and} \bibinfo{person}{Marco Costa}} (Eds.). \bibinfo{address}{Bologna, Italy}, \bibinfo{pages}{1507--1512}.
\newblock


\bibitem[Omori et~al\mbox{.}(1996)]%
        {Omori1996SingingPowerRatio}
\bibfield{author}{\bibinfo{person}{Koichi Omori}, \bibinfo{person}{Ashutosh Kacker}, \bibinfo{person}{Linda~M. Carroll}, \bibinfo{person}{William~D. Riley}, {and} \bibinfo{person}{Stanley~M. Blaugrund}.} \bibinfo{year}{1996}\natexlab{}.
\newblock \showarticletitle{Singing power ratio: quantitative evaluation of singing voice quality}.
\newblock \bibinfo{journal}{\emph{Journal of Voice}} \bibinfo{volume}{10}, \bibinfo{number}{3} (\bibinfo{year}{1996}), \bibinfo{pages}{228--235}.
\newblock
\href{https://doi.org/10.1016/s0892-1997(96)80003-8}{doi:\nolinkurl{10.1016/s0892-1997(96)80003-8}}


\bibitem[Radford et~al\mbox{.}(2023)]%
        {Whisper}
\bibfield{author}{\bibinfo{person}{Alec Radford}, \bibinfo{person}{Jong~Wook Kim}, \bibinfo{person}{Tao Xu}, \bibinfo{person}{Greg Brockman}, \bibinfo{person}{Christine McLeavey}, {and} \bibinfo{person}{Ilya Sutskever}.} \bibinfo{year}{2023}\natexlab{}.
\newblock \showarticletitle{Robust Speech Recognition via Large-Scale Weak Supervision}. In \bibinfo{booktitle}{\emph{Proceedings of the 40th International Conference on Machine Learning, ICML 2023, 23-29 July 2023, Honolulu, Hawaii, USA}} \emph{(\bibinfo{series}{Proceedings of Machine Learning Research}, Vol.~\bibinfo{volume}{202})}. \bibinfo{publisher}{PMLR}, \bibinfo{pages}{28492--28518}.
\newblock


\bibitem[Snyder et~al\mbox{.}(2018)]%
        {Snyder2018Xvectors}
\bibfield{author}{\bibinfo{person}{David Snyder}, \bibinfo{person}{Daniel Garcia-Romero}, \bibinfo{person}{Gregory Sell}, \bibinfo{person}{Daniel Povey}, {and} \bibinfo{person}{Sanjeev Khudanpur}.} \bibinfo{year}{2018}\natexlab{}.
\newblock \showarticletitle{X-vectors: Robust {DNN} Embeddings for Speaker Recognition}. In \bibinfo{booktitle}{\emph{2018 IEEE International Conference on Acoustics, Speech and Signal Processing (ICASSP)}}. \bibinfo{pages}{5329--5333}.
\newblock
\href{https://doi.org/10.1109/ICASSP.2018.8461375}{doi:\nolinkurl{10.1109/ICASSP.2018.8461375}}


\bibitem[Sun et~al\mbox{.}(2023)]%
        {Sun2023TGCritic}
\bibfield{author}{\bibinfo{person}{Xiaoheng Sun}, \bibinfo{person}{Yuejie Gao}, \bibinfo{person}{Hanyao Lin}, {and} \bibinfo{person}{Huaping Liu}.} \bibinfo{year}{2023}\natexlab{}.
\newblock \showarticletitle{{TG-Critic}: {A} Timbre-Guided Model For Reference-Independent Singing Evaluation}. In \bibinfo{booktitle}{\emph{2023 IEEE International Conference on Acoustics, Speech and Signal Processing (ICASSP)}}. \bibinfo{pages}{1--5}.
\newblock
\href{https://doi.org/10.1109/ICASSP49357.2023.10096309}{doi:\nolinkurl{10.1109/ICASSP49357.2023.10096309}}


\bibitem[Sundberg(1974)]%
        {sundberg1974articulatory}
\bibfield{author}{\bibinfo{person}{Johan Sundberg}.} \bibinfo{year}{1974}\natexlab{}.
\newblock \showarticletitle{Articulatory interpretation of the 'singing formant'}.
\newblock \bibinfo{journal}{\emph{The Journal of the Acoustical Society of America}} \bibinfo{volume}{55}, \bibinfo{number}{4} (\bibinfo{year}{1974}), \bibinfo{pages}{838--844}.
\newblock
\href{https://doi.org/10.1121/1.1914609}{doi:\nolinkurl{10.1121/1.1914609}}


\bibitem[Tsai and Lee(2011)]%
        {Tsai2011Karaoke}
\bibfield{author}{\bibinfo{person}{Wei-Ho Tsai} {and} \bibinfo{person}{Hsin-Chieh Lee}.} \bibinfo{year}{2011}\natexlab{}.
\newblock \showarticletitle{An automated singing evaluation method for karaoke systems}. In \bibinfo{booktitle}{\emph{2011 IEEE International Conference on Acoustics, Speech and Signal Processing (ICASSP)}}. \bibinfo{pages}{2428--2431}.
\newblock
\href{https://doi.org/10.1109/ICASSP.2011.5946032}{doi:\nolinkurl{10.1109/ICASSP.2011.5946032}}


\bibitem[Tsai and Lee(2012)]%
        {Tsai2012AutomaticEvaluationFeatures}
\bibfield{author}{\bibinfo{person}{Wei-Ho Tsai} {and} \bibinfo{person}{Hsin-Chieh Lee}.} \bibinfo{year}{2012}\natexlab{}.
\newblock \showarticletitle{Automatic evaluation of karaoke singing based on pitch, volume, and rhythm features}.
\newblock \bibinfo{journal}{\emph{IEEE Transactions on Audio, Speech, and Language Processing}} \bibinfo{volume}{20}, \bibinfo{number}{4} (\bibinfo{year}{2012}), \bibinfo{pages}{1233--1243}.
\newblock
\href{https://doi.org/10.1109/TASL.2011.2174200}{doi:\nolinkurl{10.1109/TASL.2011.2174200}}


\bibitem[Vaswani et~al\mbox{.}(2017)]%
        {Vaswani2017Attention}
\bibfield{author}{\bibinfo{person}{Ashish Vaswani}, \bibinfo{person}{Noam Shazeer}, \bibinfo{person}{Niki Parmar}, \bibinfo{person}{Jakob Uszkoreit}, \bibinfo{person}{Llion Jones}, \bibinfo{person}{Aidan~N Gomez}, \bibinfo{person}{{\L}ukasz Kaiser}, {and} \bibinfo{person}{Illia Polosukhin}.} \bibinfo{year}{2017}\natexlab{}.
\newblock \showarticletitle{Attention is all you need}.
\newblock \bibinfo{journal}{\emph{Advances in neural information processing systems}}  \bibinfo{volume}{30} (\bibinfo{year}{2017}).
\newblock


\bibitem[Wang et~al\mbox{.}(2023)]%
        {WeSpeaker}
\bibfield{author}{\bibinfo{person}{Hongji Wang}, \bibinfo{person}{Chengdong Liang}, \bibinfo{person}{Shuai Wang}, \bibinfo{person}{Zhengyang Chen}, \bibinfo{person}{Binbin Zhang}, \bibinfo{person}{Xu Xiang}, \bibinfo{person}{Yanlei Deng}, {and} \bibinfo{person}{Yanmin Qian}.} \bibinfo{year}{2023}\natexlab{}.
\newblock \showarticletitle{Wespeaker: A research and production oriented speaker embedding learning toolkit}. In \bibinfo{booktitle}{\emph{ICASSP 2023-2023 IEEE International Conference on Acoustics, Speech and Signal Processing (ICASSP)}}. IEEE, \bibinfo{pages}{1--5}.
\newblock


\bibitem[Wang et~al\mbox{.}(2024a)]%
        {wang2024muchinarxiv}
\bibfield{author}{\bibinfo{person}{Zihao Wang}, \bibinfo{person}{Shuyu Li}, \bibinfo{person}{Tao Zhang}, \bibinfo{person}{Qi Wang}, \bibinfo{person}{Pengfei Yu}, \bibinfo{person}{Jinyang Luo}, \bibinfo{person}{Yan Liu}, \bibinfo{person}{Ming Xi}, {and} \bibinfo{person}{Kejun Zhang}.} \bibinfo{year}{2024}\natexlab{a}.
\newblock \bibinfo{title}{MuChin: A Chinese Colloquial Description Benchmark for Evaluating Language Models in the Field of Music}.
\newblock
\showeprint[arxiv]{2402.09871}~[cs.SD]
\urldef\tempurl%
\url{https://arxiv.org/abs/2402.09871}
\showURL{%
\tempurl}


\bibitem[Wang et~al\mbox{.}(2024b)]%
        {wang2024muchin}
\bibfield{author}{\bibinfo{person}{Zihao Wang}, \bibinfo{person}{Shuyu Li}, \bibinfo{person}{Tao Zhang}, \bibinfo{person}{Qi Wang}, \bibinfo{person}{Pengfei Yu}, \bibinfo{person}{Jinyang Luo}, \bibinfo{person}{Yan Liu}, \bibinfo{person}{Ming Xi}, {and} \bibinfo{person}{Kejun Zhang}.} \bibinfo{year}{2024}\natexlab{b}.
\newblock \showarticletitle{MuChin: A Chinese Colloquial Description Benchmark for Evaluating Language Models in the Field of Music}. In \bibinfo{booktitle}{\emph{Proceedings of the Thirty-Third International Joint Conference on Artificial Intelligence, {IJCAI-24}}}, \bibfield{editor}{\bibinfo{person}{Kate Larson}} (Ed.). \bibinfo{publisher}{International Joint Conferences on Artificial Intelligence Organization}, \bibinfo{pages}{7771--7779}.
\newblock
\href{https://doi.org/10.24963/ijcai.2024/860}{doi:\nolinkurl{10.24963/ijcai.2024/860}}
\newblock
\shownote{AI, Arts \& Creativity}.


\bibitem[Wang et~al\mbox{.}(2024c)]%
        {2024muditarxiv}
\bibfield{author}{\bibinfo{person}{Zihao Wang}, \bibinfo{person}{Haoxuan Liu}, \bibinfo{person}{Jiaxing Yu}, \bibinfo{person}{Tao Zhang}, \bibinfo{person}{Yan Liu}, {and} \bibinfo{person}{Kejun Zhang}.} \bibinfo{year}{2024}\natexlab{c}.
\newblock \showarticletitle{MuDiT \& MuSiT: Alignment with Colloquial Expression in Description-to-Song Generation}.
\newblock \bibinfo{journal}{\emph{arXiv preprint arXiv:2407.03188}} (\bibinfo{year}{2024}).
\newblock


\bibitem[Wang et~al\mbox{.}(2024d)]%
        {wang2024samoyearxiv}
\bibfield{author}{\bibinfo{person}{Zihao Wang}, \bibinfo{person}{Le Ma}, \bibinfo{person}{Yongsheng Feng}, \bibinfo{person}{Xin Pan}, \bibinfo{person}{Yuhang Jin}, {and} \bibinfo{person}{Kejun Zhang}.} \bibinfo{year}{2024}\natexlab{d}.
\newblock \bibinfo{title}{SaMoye: Zero-shot Singing Voice Conversion Model Based on Feature Disentanglement and Enhancement}.
\newblock
\showeprint[arxiv]{2407.07728}~[cs.SD]
\urldef\tempurl%
\url{https://arxiv.org/abs/2407.07728}
\showURL{%
\tempurl}


\bibitem[Wang et~al\mbox{.}(2024e)]%
        {wang2024remastarxiv}
\bibfield{author}{\bibinfo{person}{Zihao Wang}, \bibinfo{person}{Le Ma}, \bibinfo{person}{Chen Zhang}, \bibinfo{person}{Bo Han}, \bibinfo{person}{Yunfei Xu}, \bibinfo{person}{Yikai Wang}, \bibinfo{person}{Xinyi Chen}, \bibinfo{person}{HaoRong Hong}, \bibinfo{person}{Wenbo Liu}, \bibinfo{person}{Xinda Wu}, {and} \bibinfo{person}{Kejun Zhang}.} \bibinfo{year}{2024}\natexlab{e}.
\newblock \bibinfo{title}{REMAST: Real-time Emotion-based Music Arrangement with Soft Transition}.
\newblock
\showeprint[arxiv]{2305.08029}~[cs.SD]
\urldef\tempurl%
\url{https://arxiv.org/abs/2305.08029}
\showURL{%
\tempurl}


\bibitem[Wang et~al\mbox{.}(2025a)]%
        {remast}
\bibfield{author}{\bibinfo{person}{Zihao Wang}, \bibinfo{person}{Le Ma}, \bibinfo{person}{Chen Zhang}, \bibinfo{person}{Bo Han}, \bibinfo{person}{Yunfei Xu}, \bibinfo{person}{Yikai Wang}, \bibinfo{person}{Xinyi Chen}, \bibinfo{person}{Haorong Hong}, \bibinfo{person}{Wenbo Liu}, \bibinfo{person}{Xinda Wu}, {and} \bibinfo{person}{Kejun Zhang}.} \bibinfo{year}{2025}\natexlab{a}.
\newblock \showarticletitle{REMAST: Real-Time Emotion-Based Music Arrangement With Soft Transition}.
\newblock \bibinfo{journal}{\emph{IEEE Transactions on Affective Computing}} \bibinfo{volume}{16}, \bibinfo{number}{2} (\bibinfo{year}{2025}), \bibinfo{pages}{1016--1030}.
\newblock
\href{https://doi.org/10.1109/TAFFC.2024.3486224}{doi:\nolinkurl{10.1109/TAFFC.2024.3486224}}


\bibitem[Wang et~al\mbox{.}(2025b)]%
        {mudit-acl}
\bibfield{author}{\bibinfo{person}{Zihao Wang}, \bibinfo{person}{Jiaxing Yu}, \bibinfo{person}{Haoxuan Liu}, \bibinfo{person}{Zehui Zheng}, \bibinfo{person}{Yuhang Jin}, \bibinfo{person}{Shuyu Li}, \bibinfo{person}{Shulei Ji}, {and} \bibinfo{person}{Kejun Zhang}.} \bibinfo{year}{2025}\natexlab{b}.
\newblock \showarticletitle{Generative Music Models' Alignment with Professional and Amateur Users' Expectations}. In \bibinfo{booktitle}{\emph{Findings of the Association for Computational Linguistics: ACL 2025}}, \bibfield{editor}{\bibinfo{person}{Wanxiang Che}, \bibinfo{person}{Joyce Nabende}, \bibinfo{person}{Ekaterina Shutova}, {and} \bibinfo{person}{Mohammad~Taher Pilehvar}} (Eds.). \bibinfo{publisher}{Association for Computational Linguistics}, \bibinfo{address}{Vienna, Austria}, \bibinfo{pages}{6909--6920}.
\newblock
\showISBNx{979-8-89176-256-5}
\href{https://doi.org/10.18653/v1/2025.findings-acl.360}{doi:\nolinkurl{10.18653/v1/2025.findings-acl.360}}


\bibitem[Wang et~al\mbox{.}(2022a)]%
        {2022songdriverarxiv}
\bibfield{author}{\bibinfo{person}{Zihao Wang}, \bibinfo{person}{Kejun Zhang}, \bibinfo{person}{Yuxing Wang}, \bibinfo{person}{Chen Zhang}, \bibinfo{person}{Qihao Liang}, \bibinfo{person}{Pengfei Yu}, \bibinfo{person}{Yongsheng Feng}, \bibinfo{person}{Wenbo Liu}, \bibinfo{person}{Yikai Wang}, \bibinfo{person}{Yuntao Bao}, {and} \bibinfo{person}{Yiheng Yang}.} \bibinfo{year}{2022}\natexlab{a}.
\newblock \showarticletitle{SongDriver: Real-time Music Accompaniment Generation without Logical Latency nor Exposure Bias}.
\newblock \bibinfo{journal}{\emph{arXiv preprint arXiv:2209.06054}} (\bibinfo{year}{2022}).
\newblock


\bibitem[Wang et~al\mbox{.}(2022b)]%
        {2022songdriver}
\bibfield{author}{\bibinfo{person}{Zihao Wang}, \bibinfo{person}{Kejun Zhang}, \bibinfo{person}{Yuxing Wang}, \bibinfo{person}{Chen Zhang}, \bibinfo{person}{Qihao Liang}, \bibinfo{person}{Pengfei Yu}, \bibinfo{person}{Yongsheng Feng}, \bibinfo{person}{Wenbo Liu}, \bibinfo{person}{Yikai Wang}, \bibinfo{person}{Yuntao Bao}, {and} \bibinfo{person}{Yiheng Yang}.} \bibinfo{year}{2022}\natexlab{b}.
\newblock \showarticletitle{SongDriver: Real-time Music Accompaniment Generation without Logical Latency nor Exposure Bias}. In \bibinfo{booktitle}{\emph{Proceedings of the 30th ACM International Conference on Multimedia}} (Lisboa, Portugal) \emph{(\bibinfo{series}{MM '22})}. \bibinfo{publisher}{Association for Computing Machinery}, \bibinfo{address}{New York, NY, USA}, \bibinfo{pages}{1057–1067}.
\newblock
\showISBNx{9781450392037}
\href{https://doi.org/10.1145/3503161.3548368}{doi:\nolinkurl{10.1145/3503161.3548368}}


\bibitem[Wapnick et~al\mbox{.}(1997)]%
        {Wapnick1997ExpertConsensus}
\bibfield{author}{\bibinfo{person}{Joel Wapnick}, \bibinfo{person}{Elizabeth Ekholm}, {and} \bibinfo{person}{John D'Ombrain}.} \bibinfo{year}{1997}\natexlab{}.
\newblock \showarticletitle{Expert consensus in solo voice performance evaluation}.
\newblock \bibinfo{journal}{\emph{Journal of Voice}} \bibinfo{volume}{11}, \bibinfo{number}{4} (\bibinfo{year}{1997}), \bibinfo{pages}{429--436}.
\newblock
\href{https://doi.org/10.1016/s0892-1997(97)80039-2}{doi:\nolinkurl{10.1016/s0892-1997(97)80039-2}}


\bibitem[Wu et~al\mbox{.}(2023)]%
        {wu2023melodyglm}
\bibfield{author}{\bibinfo{person}{Xinda Wu}, \bibinfo{person}{Zhijie Huang}, \bibinfo{person}{Kejun Zhang}, \bibinfo{person}{Jiaxing Yu}, \bibinfo{person}{Xu Tan}, \bibinfo{person}{Tieyao Zhang}, \bibinfo{person}{Zihao Wang}, {and} \bibinfo{person}{Lingyun Sun}.} \bibinfo{year}{2023}\natexlab{}.
\newblock \showarticletitle{MelodyGLM: multi-task pre-training for symbolic melody generation}.
\newblock \bibinfo{journal}{\emph{arXiv preprint arXiv:2309.10738}} (\bibinfo{year}{2023}).
\newblock


\bibitem[Yuan et~al\mbox{.}(2025)]%
        {yuan2025yue}
\bibfield{author}{\bibinfo{person}{Ruibin Yuan}, \bibinfo{person}{Hanfeng Lin}, \bibinfo{person}{Shawn Guo}, \bibinfo{person}{Ge Zhang}, \bibinfo{person}{Jiahao Pan}, \bibinfo{person}{Yongyi Zang}, \bibinfo{person}{Haohe Liu}, \bibinfo{person}{Xingjian Du}, \bibinfo{person}{Xeron Du}, \bibinfo{person}{Zhen Ye}, \bibinfo{person}{Tianyu Zheng}, \bibinfo{person}{Yinghao Ma}, \bibinfo{person}{Minghao Liu}, \bibinfo{person}{Lijun Yu}, \bibinfo{person}{Zeyue Tian}, \bibinfo{person}{Ziya Zhou}, \bibinfo{person}{Liumeng Xue}, \bibinfo{person}{Xingwei Qu}, \bibinfo{person}{Yizhi Li}, \bibinfo{person}{Tianhao Shen}, \bibinfo{person}{Ziyang Ma}, \bibinfo{person}{Shangda Wu}, \bibinfo{person}{Jun Zhan}, \bibinfo{person}{Chunhui Wang}, \bibinfo{person}{Yatian Wang}, \bibinfo{person}{Xiaohuan Zhou}, \bibinfo{person}{Xiaowei Chi}, \bibinfo{person}{Xinyue Zhang}, \bibinfo{person}{Zhenzhu Yang}, \bibinfo{person}{Yiming Liang}, \bibinfo{person}{Xiangzhou Wang}, \bibinfo{person}{Shansong Liu}, \bibinfo{person}{Lingrui Mei},
  \bibinfo{person}{Peng Li}, \bibinfo{person}{Yong Chen}, \bibinfo{person}{Chenghua Lin}, \bibinfo{person}{Xie Chen}, \bibinfo{person}{Gus Xia}, \bibinfo{person}{Zhaoxiang Zhang}, \bibinfo{person}{Chao Zhang}, \bibinfo{person}{Wenhu Chen}, \bibinfo{person}{Xinyu Zhou}, \bibinfo{person}{Xipeng Qiu}, \bibinfo{person}{Roger Dannenberg}, \bibinfo{person}{Jiaheng Liu}, \bibinfo{person}{Jian Yang}, \bibinfo{person}{Stephen Huang}, \bibinfo{person}{Wei Xue}, \bibinfo{person}{Xu Tan}, {and} \bibinfo{person}{Yike Guo}.} \bibinfo{year}{2025}\natexlab{}.
\newblock \bibinfo{title}{YuE: Open Music Foundation Models for Full-Song Generation}.
\newblock \bibinfo{howpublished}{\url{https://github.com/multimodal-art-projection/YuE}}.
\newblock
\newblock
\shownote{GitHub repository}.


\bibitem[Zhang et~al\mbox{.}(2019)]%
        {Zhang2019BiDense}
\bibfield{author}{\bibinfo{person}{Ning Zhang}, \bibinfo{person}{Tao Jiang}, \bibinfo{person}{Feng Deng}, {and} \bibinfo{person}{Yan Li}.} \bibinfo{year}{2019}\natexlab{}.
\newblock \showarticletitle{Automatic singing evaluation without reference melody using bi-dense neural network}. In \bibinfo{booktitle}{\emph{ICASSP 2019 - 2019 IEEE International Conference on Acoustics, Speech and Signal Processing (ICASSP)}}. \bibinfo{pages}{466--470}.
\newblock
\href{https://doi.org/10.1109/ICASSP.2019.8683702}{doi:\nolinkurl{10.1109/ICASSP.2019.8683702}}


\end{thebibliography}

\end{document}